\documentclass[prl,aps,twocolumn,twoside,showpacs,amsmath,amssymb]{revtex4}

\usepackage{graphicx}% Include figure files
\newcommand{\sigmarms}{\sigma_\mathrm{rms}}

\newcommand{\ba}{{\bf a}}

\newcommand{\bu}{{\bf u}}
\newcommand{\bw}{{\bf w}}
\newcommand{\be}{{\bf e}}

\newcommand{\bB}{{\bf B}}
\newcommand{\bF}{{\bf F}}

\newcommand{\bK}{{\bf K}}

\newcommand{\cH}{{\cal H}}

\newcommand{\half}{\frac{1}{2}}
\newcommand{\f}{\frac}

\newcommand{\Eq}[1]{Eq.~(\ref{#1})}
\newcommand{\Fig}[1]{Fig.~\ref{#1}}
\newcommand{\Ref}[1]{(\ref{#1})}
\newcommand{\Table}[1]{Table~\ref{#1}}

\newcommand{\av}[1]{\left<{#1}\right>}
\newcommand{\disav}[1]{\left[{#1}\right]_\mathrm{av}}

\begin{document}

%\preprint{}

\title{Amorphous vortex glass phase in strongly disordered superconductors}

\author{Jack Lidmar}
\email{jlidmar@kth.se}
\affiliation{
Department of Physics,
Royal Institute of Technology, AlbaNova,
SE-106 91 Stockholm, Sweden
}

\date{December 18, 2002}

\begin{abstract}
We introduce a model describing vortices in strongly disordered
three-dimensional superconductors.  The model focuses on the
topological defects, i.e., dislocation lines, in an elastic
description of the vortex lattice.  The model is studied using Monte
Carlo simulations, revealing a glass phase at low temperatures,
separated by a continuous phase transition to the high temperature
resistive vortex liquid phase.  The critical exponents $\nu \approx
1.3$ and $\eta \approx -0.4$ characterizing the transition are
obtained from finite size scaling.
\end{abstract}

\pacs{74.60.-w,05.70.Fh,75.40.Mg}

%\keywords{} % Use showkeys class option if keyword display desired

\maketitle

%Introduction

Disorder has a profound impact on the phase diagram of type-II
superconductors.  The long range order of the Abrikosov vortex lattice
in very clean samples is destabilized upon the introduction of
disorder~\cite{Larkin} and a glass state can appear
instead~\cite{M.P.A.Fisher,FFH}.  The effects of weak disorder are
relatively well understood and may be treated within an elastic theory
of the vortex
lattice~\cite{Feigelman-89-Nattermann,Giamarchi-Doussal,D.Fisher}.
The resulting phase, known as the Bragg glass or the {\em elastic}
vortex glass, is characterized by an algebraic decay of translational
correlations and perfect orientational order, and the absence of large
dislocation loops~\cite{D.Fisher}.

For stronger disorder dislocations start to become energetically
favorable and begin to proliferate.  What happens then is still a
matter of controversy.  One possibility is that any kind of order in
the system is lost and the system becomes thermodynamically equivalent
to a vortex line liquid (although perhaps a very viscous one) with a
non-zero resistivity~\cite{FFH}.  The other possibility is that the
dislocations themselves remain pinned and the system freezes into a
different genuine thermodynamic {\em amorphous} glass phase, full of
dislocation loops, and with zero linear
resistivity.
Such a disorder driven transition from the Bragg glass to the vortex
glass have been proposed to occur as the magnetic field is
increased~\cite{Giamarchi-Doussal,Gingras-Huse}.

Unfortunately, the theoretical methods based on pinning of elastic
manifolds in random media, that were so successful for the Bragg
glass, does not easily extend to incorporate the effects of
dislocations.  In this Letter we introduce and study a model that goes
beyond a purely elastic description by including dislocations, and is
well suited for numerical simulations.

% Experiments:

Experimentally, the evidence for a vortex glass comes mainly from the
scaling relations obeyed by the current-voltage characteristics and
resistivity at a continuous phase transition~\cite{FFH}.  The first
order vortex melting line, observed in very clean samples at low
magnetic fields, is replaced by a smooth continuous transition upon
increasing the field or disorder.  The universal critical exponents
extracted from measurements in this regime provide an important
characterization of the glass transition that can be compared with
theoretical predictions.  A large number of studies have thus been
able to obtain good scaling behavior in different materials, and this
fact has been taken to support the existence of an amorphous glass
phase~\cite{vortexglass-experiment}.  This interpretation has,
however, been questioned recently~\cite{Strachan}.  In practice it may
be extremely difficult to discriminate between a true thermodynamic
glass and a very slowly moving non-equilibrium viscous vortex liquid
(as window glass), with a small, but non-zero, resistivity due to
thermally activated flux creep.

% Gauge glass:

On the theoretical side a lot of interest has focused on the so called
gauge glass model~\cite{gauge-glass}.  This approach, which is
inspired by spin glasses in random magnets, combines the $U(1)$
symmetry of a superconducting order parameter with quenched disorder
and frustration, which are believed necessary to create a glass phase,
but has no other connection with the microscopics.  The model, which
consists of an $XY$-model in a quenched random gauge field, has a
glass phase at low temperatures in three dimensions~\cite{gauge-glass}
(but probably not in two).  Once screening of the vortex interaction,
due to gauge field fluctuations, is accounted for, the glass phase
disappears~\cite{gauge-glass-screening}.  However, the way disorder
enters the model, via quenched random fluxes instead of a random core
energy coupling to the vortices, is not very realistic.  More
importantly, perhaps, the model lacks a length scale, the inter-vortex
distance, set by the external magnetic field.

% Realistic models

Attempts to study more realistic models have come to contradictory
conclusions.  Simulations of frustrated 3D $XY$-models with quenched
disorder show evidence for a direct disorder driven transition from
the Bragg glass to a vortex liquid at low disorder and fillings, and
find no clear signs of any amorphous glass phase~\cite{Olsson-Teitel}
(See, however, Ref.~\onlinecite{Kawamura}).
On the other hand recent simulations~\cite{Vestergren} of a random
pinning vortex model for much stronger disorder and very high field,
corresponding to a half-filled system, do find a continuous
transition, but with critical exponents quite far from experiments.
None of these models included screening.
Including screening in a similar model appears to destroy the glass
phase~\cite{Kawamura}, just as in the gauge glass.
Langevin simulations of vortex lines also found no glass
phase~\cite{molasses}.
Recently, Kierfeld {\em et al.}~\cite{Kierfeld} invoked a Landau
theory in terms of the dislocation density, assuming the dislocation
lines were {\em directed}, and proposed a phase diagram containing an
amorphous glass phase.
Some support for a glass phase has also been found using the boson
analogy~\cite{Kihong}.

% Model

The approach taken in this Letter starts from an elastic theory of the
vortex lattice, but modified to include dislocations.  The basic
degrees of freedom in our model is the two-dimensional displacement
vector $\bu_i$ and a tensor field $\bB$ that accounts for the
dislocations.  Since the clean Abrikosov vortex lattice usually has
hexagonal symmetry we define the model on a stacked triangular lattice
with $N=L^3$ sites and unit lattice constant, and with the
displacements $\bu_i$ living on the vertices and the field $\bB_{ij}$
living on the links connecting nearest neighbors $i$ and $j$.  The
Hamiltonian reads
\begin{equation}						\label{eq:H}
  \cH = \half \sum_{\left<ij\right>} \left(\bw_{ij} - \bF_{ij}\right)^2,
\end{equation}
where $\left<ij\right>$ denote all nearest neighbor pairs on the
stacked triangular lattice, $\bw_{ij}= \bu_i - \bu_j + \bB_{ij}$, and
$\bu_i \in \mathbb{R}^2$.  The $\bB_{ij}$ are integer multiples of the
elementary triangular lattice vectors $\ba_1 = (1,0)$, $\ba_2 =
(\half, \f{\sqrt{3}}{2})$ in the plane, i.e., $\bB_{ij} = m \ba_1 + n
\ba_2$, $m,n \in \mathbb{Z}$, and contain all the information about
the dislocations.  More precisely the dislocation density is given by
the lattice curl of $\bB$, implying that dislocations form lines with
conserved Burgers vector as they should.
A dual representation in terms of interacting dislocation lines can be
obtained by integrating out $\bu$, but \Eq{eq:H} turns out to be more
convenient for our purposes.
%
% Disorder coupling:
The coupling to disorder occurs via the random variable $\bF_{ij}$
\footnote{This corresponds to a random stress.  Other ways of
incorporating disorder would also be interesting to consider.}.
For simplicity we take the maximum possible disorder strength, with
$\bF_{ij}$ independently and uniformly distributed in the planar unit
cell of the triangular lattice.

In the absence of dislocations all $\bB_{ij}$ are zero and the model
reduces to an ordinary elastic model of the vortex lattice with
compression modulus $c_{11} = 3/2$, tilt modulus $c_{44} = 1$, and
shear modulus $c_{66} = 3/2$
\footnote{
To avoid factors of $\sqrt{3}/2$ extensive quantities are normalized
using $N$ instead of the volume $\Omega = N\sqrt{3}/2$.
}.
The generalization to other values of these (bare) elastic constants
is straightforward via the addition of a term proportional to
$\sum_{\left<ij\right>} \left( \bw_{ij} \cdot \be_{ij} \right)^2$
[$\be_{ij}$ being the unit vector connecting nearest neighbor sites
$i$ and $j$] to \Eq{eq:H} and/or by changing the coupling in the
$z$-direction to make it anisotropic.  In this work we will, however,
restrict to the isotropic model defined above.

The presence of thermally excited dislocations will renormalize the
elastic constants and eventually cause a transition at which they drop
to zero.  The renormalized elastic constants can be obtained by taking
the second derivative of the free energy with respect to a
deformation, $\bw_{ij} \to \bw_{ij} + \varepsilon_{\mu\nu} (\be_{ij}
\cdot \be_\mu) \be_\nu$.
This gives
\begin{eqnarray}					\label{eq:cijij}
C_{\mu\nu | \rho\sigma} &=&
\f{\partial^2 f}{\partial\varepsilon_{\mu\nu}
               \partial\varepsilon_{\rho\sigma}} =
\f{1}{2}\delta_{\mu\rho}\delta_{\nu\sigma}
\left(3-\delta_{\mu,\hat z}\right)  \nonumber \\
&& -
   \f{N}{T} \left( 
\left< \sigma_{\mu\nu} \sigma_{\rho\sigma} \right>
 - \left< \sigma_{\mu\nu} \right> \left< \sigma_{\rho\sigma} \right>
\right),
\end{eqnarray}
where $\left<\cdots\right>$ denotes a thermal average, and the
$3 \times 2$ stress tensor $\sigma$ is given by
\begin{equation}					\label{eq:stress}
  \sigma_{\mu\nu}
   =	\f{1}{N} \sum_{\left<ij\right>}
	\left( \be_{ij} \cdot \be_\mu \right)
	\left( \be_\nu \cdot \left[ \bB_{ij} - \bF_{ij}
	\right] \right).
\end{equation}
Unfortunately, both of these vanish upon averaging over the disorder
realizations, $\left[\left<\sigma\right>\right]_\mathrm{av} =
\left[C\right]_\mathrm{av} = 0$, since $\disav{f}$ becomes independent
of $\varepsilon_{\mu\nu}$ for the fully disordered model, where the
averaging includes all possible deformations.  A non-zero
$\left<\sigma\right>$ is still indicative of a phase with rigidity,
and hence we use
\begin{equation}					\label{eq:sigmarms}
  \sigma_\mathrm{rms}^2 =
\disav{\sum_{\mu\nu} {\left< \sigma_{\mu\nu} \right>^2}}
= \disav{ \sum_{\mu\nu} {\left< \sigma_{\mu\nu} \right>}_\alpha {\left<
	\sigma_{\mu\nu} \right>}_\beta }
\end{equation}
instead, where the second form, which involves two different replicas
$\alpha$ and $\beta$ to avoid any bias in the expectation value, is
the one used in the actual simulations.
Alternatively, $\sigmarms^2$ can be calculated from \Eq{eq:cijij}
using that $\left[C\right]_\mathrm{av} = 0$,
\begin{equation}					\label{eq:sigmarmsalt}
  \sigmarms^2 = \disav{ \sum_{\mu\nu} \av{ \sigma_{\mu\nu}^2 } }
		- \f{8 T}{N}.
\end{equation}
In the solid low temperature phase (if it exists) $\sigmarms^2$ will
acquire a non-zero expectation value, while in the liquid it
approaches zero.  Close to a continuous phase transition it obeys the
finite size scaling form,
\begin{equation}				\label{eq:sigma-scaling}
  \sigma_\mathrm{rms} = L^{1-d} \hat\sigma(t L^{1/\nu}),
\end{equation}
where $t = (T-T_c)/T_c$ is the reduced temperature, $\nu$ the
correlation length exponent, $d = 3$ the dimensionality, and
$\hat\sigma$ is a scaling function
\footnote{
The absence of any nontrivial anomalous dimension in
\Eq{eq:sigma-scaling} is a result of the underlying periodicity in the
model.  At a fixed point with such periodicity the scaling dimension
of $\bu$ must vanish, and hence $\bw$ and $\bB$ scales as $L^{-1}$ and
$\sigma$ as $L^{1-d}$.}.

Lattice order may be probed by the structure function $S_\bK =
\f{1}{N} \left< \left| \sum_i e^{i\bK \cdot \bu_i} \right|^2 \right>$
at a reciprocal lattice vector $\bK$ of the triangular lattice, but
for strong disorder, when the system might freeze into a highly
disordered glassy state, this is not useful.
To quantify order in a glass one may instead consider an overlap order
parameter,
\begin{equation}					\label{eq:overlap}
  q_\bK = \f{1}{N} \sum_i 
    e^{i \bK \cdot \left(\bu_i^\alpha - \bu_i^\beta \right)},
\end{equation}
between two replicas $\alpha$ and $\beta$.
The order parameter susceptibility, given by
\begin{equation}					\label{eq:chi}
\chi_\bK = {N} \left[\left< \left| q_\bK \right|^2 \right>\right]_\mathrm{av},
\end{equation}
obeys the finite size scaling relation
\begin{equation}					\label{eq:chi-scaling}
  \chi_\bK = L^{2-\eta} \hat \chi(t L^{1/\nu}),
\end{equation}
which defines the critical exponent $\eta$.

We study this model via Monte Carlo (MC) simulations.  One MC sweep
consists of $N$ randomly chosen heat bath updates of the $\bu_i$ and
$12N$ Metropolis updates of $\bB_{ij}$, where $N = L^3$ is the number
of sites of the lattice.  In order to overcome the slow glassy
dynamics in the fully disordered system an exchange (or parallel
tempering) MC algorithm~\cite{exchange} is used, where several systems
with identical disorder configurations but different temperatures are
simulated in parallel and sometimes exchanged.  The temperatures were
chosen in a range around $T_c$, such that the acceptance rate of
temperature exchanges were no less than $15\%$.

\begin{figure}
\includegraphics[width=0.96\linewidth]{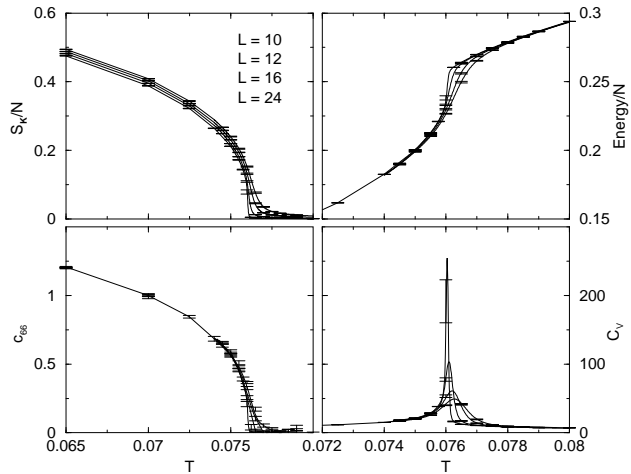}
\caption{						\label{fig:clean}
Clean system.  The solid lines are interpolations using histogram
reweighting~\cite{reweight}.  }
\end{figure}

We begin with a brief discussion of the clean model (all $\bF_{ij} =
0$).  Figure~\ref{fig:clean} shows a sharp drop in the structure
function $S_\bK$ and in the renormalized shear modulus $c_{66}$, and a
discontinuous jump in the average energy, with a corresponding
approximate delta-function peak in the specific heat $C_V$.
The very sharp variation in these quantities at $T_c \approx 0.076$
signifies a discontinuous first order transition in accordance with
experiments~\cite{first-order-melting-expr} as well as numerical
simulations of frustrated $XY$-models~\cite{Hu-1997} and vortex line
models~\cite{helicity, Nordborg-Blatter}.
This is also confirmed by looking at histograms for the energy or the
structure function, which show an increasingly growing double-peak
structure developing for the largest system sizes. 

% Disordered model.

% Test of convergence

\begin{figure}
\includegraphics[width=0.91\linewidth]{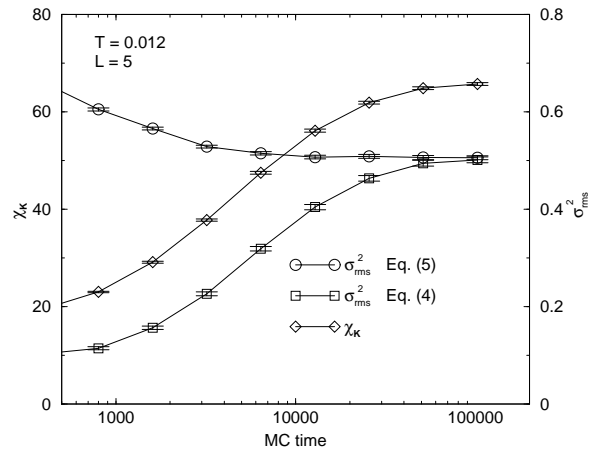}
\caption{						\label{fig:relax}
Test of equilibration in the glass.  As seen, $\sigmarms^2$ from
\Eq{eq:sigmarms} and \Ref{eq:sigmarmsalt} converge to a common value
from below and above, which provides a very useful check on the
equilibration.
}
\end{figure}

We now turn to the disordered model.  In the maximally disordered
model the dynamics becomes very slow at low temperatures, and it is
important to check the convergence of the MC simulation. To do so we
monitor several quantities averaged over increasingly longer
simulation times, after discarding an equally long time for
equilibration.  In \Fig{fig:relax} we plot, as an example,
$\sigmarms^2$ calculated from \Eq{eq:sigmarms} and
\Ref{eq:sigmarmsalt}, and $\chi_\mathbf{K}$ as a function of MC time
for the size $L = 5$ and the lowest $T$ used in the exchange MC for
this size.  As seen the curves saturate and become time independent
which indicates that the simulation has converged at around 100000
sweeps.  Similar checks are done for all sizes.  Typically, 4000--7000
disorder realizations are used to form disorder averages.

\begin{figure}
\includegraphics[width=0.85\linewidth]{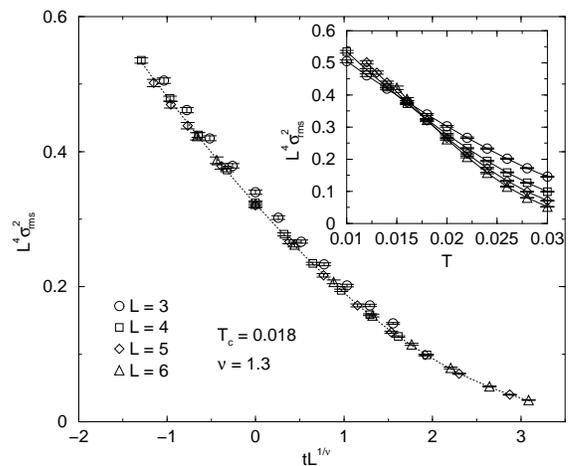}
\caption{						\label{fig:sigma}
Finite size scaling collapse of MC data for $\sigmarms^2$ according to
\Eq{eq:sigma-scaling}.  Corrections to scaling are clearly visible for
the smallest size, $L=3$, which was therefore omitted from the fitting
procedure.  The crossing point in the inset gives an estimate of
$T_c$.}
\end{figure}

\begin{figure}
\includegraphics[width=0.9\linewidth]{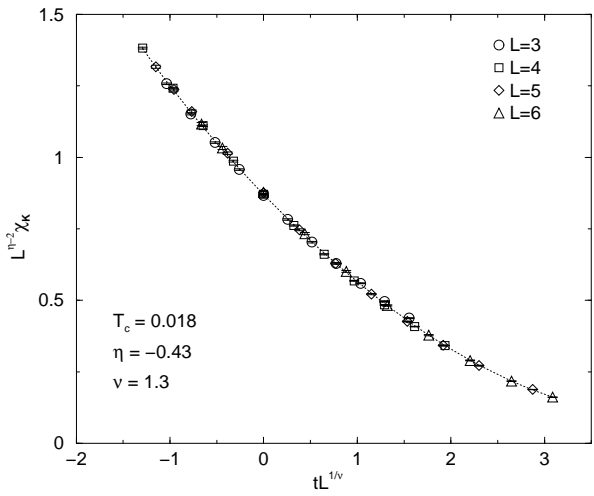}
\caption{						\label{fig:chi}
Finite size scaling collapse of MC data for $\chi_\bK$ using
\Eq{eq:chi-scaling}.
}
\end{figure}

In the inset of \Fig{fig:sigma} we show MC data for $L^4\sigmarms^2$
vs.\ $T$ for several different system sizes, which according to
\Eq{eq:sigma-scaling} should become independent of $L$ exactly at
$T=T_c$.  Indeed, the curves cross, which indicates that the model has
a glass phase below this point, with a non-zero $T_c \approx 0.018$.
The main part shows a finite size scaling collapse of the same data,
which gives an estimate of the critical exponent $\nu$.
The structure function $S_\bK$ stays constant $\approx 1$ independent of
temperature and system size, implying a total lack of crystalline
order throughout the glass phase.
Also, no signature of the transition is found in the density of
dislocations, which remains finite down to the lowest temperatures.
In \Fig{fig:chi} we plot a finite size scaling collapse using
\Eq{eq:chi-scaling} of $\chi_\bK$ (averaged over the smallest
reciprocal lattice vectors).
As a measure of the quality of the data collapses we use the error in
a $\chi^2$--fit of the scaled data to a polynomial (indicated by the
dotted lines in the figures).  The critical parameters are estimated
by adjusting them to minimize this error, and the errorbars are
estimated by varying them until the data no longer scales.
The final results for the critical parameters are summarized in
\Table{tab:exponents}.

\begin{table}[b]
\caption{						\label{tab:exponents}
Estimates of the critical properties of the model.
}
\begin{ruledtabular}
\begin{tabular}{ccc}
 $T_c$ & $\nu$ & $\eta$ \\
%\hline
 $0.018 \pm 0.002$ & $1.3 \pm 0.2$  & $-0.4 \pm 0.1$ \\
\end{tabular}
\end{ruledtabular}
\end{table}

In summary, we have introduced a model of vortex lattices subjected to
disorder.  The model goes beyond an elastic description by including
dislocations.  Such a description should be valid as soon as the
screening length is larger than the inter-vortex distance, $\lambda
\gtrsim a_v$.
The model is studied via MC simulations and finite size scaling.
Without disorder the model shows a first order melting transition.
For strong disorder, the model has a low-temperature glass phase where
the dislocation loops are pinned by the disorder.
The glass melts via a continuous transition characterized by the
critical exponents in \Table{tab:exponents}.
We propose that the model gives an effective description of the vortex
glass transition in type-II superconductors.
Experiments usually find a correlation length exponent in the range $1
< \nu < 2$~\cite{vortexglass-experiment}, which compares well with the
present value $\nu = 1.3 \pm 0.2$.
Very recent simulations on disordered 3D $XY$-models find evidence for
a glass transition with $\nu = 1.5 \pm 0.12$ or $\nu = 1.1 \pm 0.2$,
$\eta = -0.5 \pm 0.1$~\cite{Olsson-Kawamura}, consistent with the
present study.
%
% Future directions:
It would be interesting to extend the study to include dynamic
quantities, and to map out the phase diagram for intermediate disorder
strengths and other values of the bare elastic constants.

\vspace{-16pt}

%\bibliography{avg}

\end{document}